**Helical Structures in Vertically Aligned Dust Particle Chains in a Complex Plasma**

Truell W. Hyde, Jie Kong, Lorin S. Matthews

CASPER (Center for Astrophysics, Space Physics, and Engineering Research) Baylor University, Waco, Texas 76798-7310, USA

**Abstract**

Self-assembly of structures from vertically aligned, charged dust particle bundles within a glass box placed on the lower, powered electrode of a RF GEC cell were produced and examined experimentally. Self-organized formation of one-dimensional vertical chains, two-dimensional zigzag structures and three-dimensional helical structures of triangular, quadrangular, pentagonal, hexagonal, and heptagonal symmetries are shown to occur. System evolution is shown to progress from a one-dimensional chain structure, through a zigzag transition to a two-dimensional, spindle-like structure and then to various three-dimensional, helical structures exhibiting multiple symmetries. Stable configurations are found to be dependent upon the system confinement, $\gamma^2 = (\omega_{0h}/\omega_{0v})^2$ (where $\omega_{0h,v}$ are the horizontal and vertical dust resonance frequencies), the total number of particles within a bundle and the RF power. For clusters having fixed numbers of particles, the RF power at which structural transitions occur is repeatable and exhibits no observable hysteresis. The critical conditions for these structural transitions as well as the basic symmetry exhibited by the one-, two- and three-dimensional structures that subsequently develop are in good agreement with the theoretically predicted configurations of minimum energy determined employing molecular dynamics simulations for charged dust particles confined in a prolate, spheroidal potential as presented theoretically by Kamimura and Ishihara [10].

## I.  Introduction

Physicists have studied cold confined ion/electron systems for decades, beginning with the examination of particles interacting through a bare Coulomb potential by J.J. Thomson at the turn of the previous century [1]. Recently these studies have been rekindled, in part due to renewed interest in interacting particles in low dimensions and confined geometries. Such quasi-one-dimensional (Q1D), mesoscopic systems cover a wide regime of interest within physics, ranging from the examination of electrons 'floating' on the surface of liquid helium 3 or 'falling' through liquid helium 4, to the stable confinement of atomic ions within a surface-electrode trap [2, 3].

Over the past eighteen years, complex (dusty) plasmas have provided a versatile analog for studying the systems mentioned above [4, 5]. Complex plasmas consist of partially ionized gases containing micron- or nano-sized, dust particles. These particles collect ions and electrons from the plasma, in general obtaining a negative charge due to the higher mobility of the electrons, and are experimentally observable through the scattering of laser light. Dust-dust particle interactions are well described by the Debye-Huckel (Yukawa) potential and the system of dust particles plus plasma can be assumed to maintain overall charge neutrality.

Quasi-one- and two-dimensional systems can form within a complex plasma over a variety of confining and interparticle potentials. The majority of studies to date have assumed a biharmonic confinement and a

Debye-Huckel (Yukawa) interaction potential, which allows the formation of horizontal (i.e., perpendicular to the gravitational force) one- and two-dimensional symmetric structures [11, 12, 13, 14]. These have been studied extensively and shown to produce stable, symmetric configurations ranging from single chains to zigzag, spindlelike, shell, or helical structures depending on the total number of particles, the confining potential, and the shielding strength of the plasma. More recently, vertically aligned (i.e., parallel to the gravitational force) one-dimensional chains have also been examined [9].

In a recent paper by Kamimura & Ishihara [10], a molecular dynamics simulation was employed to model a complex plasma and determine the minimum energy states defined by the confinement. This simulation confirmed the transition from a one-dimensional chain to a two-dimensional zigzag structure and predicted the formation of helical structure for appropriate system conditions and confinement.

In this paper, these theoretical predictions are examined experimentally. Specifically, the transition between a vertically aligned one-dimensional particle chain to vertically aligned two- and three-dimensional bundle structures are explored. Each is examined in detail for varying numbers of particles and changing confinement potentials. In addition to showing agreement with the results provided by Kamimura & Ishihara [10], this paper also shows that higher order structural symmetries occur, which are dependent on both plasma operating parameters and system confinement.

The paper is arranged in the following manner. In Section II, a brief description of the relevant previous work is presented. In Section III, the experimental method is provided as well as a brief description of the apparatus employed. Experimental results are presented in Section IV and a discussion of the data and conclusions is given in Section V.

## II.   Previous Work

In 1998, Candido *et al.* [12] numerically examined the stable-state configuration and dynamics of a classical 2D system of charged particles assuming a harmonic confinement and a Coulombic interparticle interaction potential. It was shown that the structural symmetry of the system was related to the anisotropy of the confinement potential, α, the number of particles and the screening length, κ. It was also determined that critical parameters of κ and α existed for structural transitions. As anisotropy increased, overall system symmetry evolved from a series of circular shells to ellipses through a decrease in number of shells; this occurred through a series of structural phase transitions in which inner shells collapsed into a line. In the limit of extreme anisotropy of the confinement potential, a 1D configuration of particles was formed.

In 2005, Arp *et al.* calculated the ion-drag force within a glass box placed on the lower electrode, as derived from a simulation by Khrapak [15, 16]. They showed that the ion flow exhibited low Mach numbers (M ~ 0.2) leading to the conclusion that the ion drag force was smaller than the electric field force within the glass box by at least two orders of magnitude. This led them to the conclusion that the ion-drag force does not contribute to the topology of the trap.

In 2006, Melzer [13] experimentally examined the zigzag transition predicted theoretically by Schiffer [19] for horizontal finite dust clusters. The zigzag transition was found to be driven by an increase in particle number and/or change in the anisotropy of the confinement. Surprisingly over the parameter range investigated, zigzag transitions between one- and two-dimensional structures were not produced by changes in plasma power, which impacts both the electron and ion density in the sheath.

In 2012, Kamimura and Ishihara [10], employed a molecular dynamics simulation to determine configurations of minimum energy (CME) for a complex plasma and found them to be dependent upon the total number of particles (N), the prolateness parameter ($\alpha^{-1}$), and the plasma screening length ($\lambda$). They assumed both a radial and axial harmonic potential confinement, a Debye-Huckel interparticle interaction, overall charge neutrality, no plasma source, no gravity, no recombination and no ion wake potential. Under these conditions, they theoretically determined CME's that produced a single particle chain, the zigzag two-dimensional structure discussed above and helical strings. (See Fig 1.)

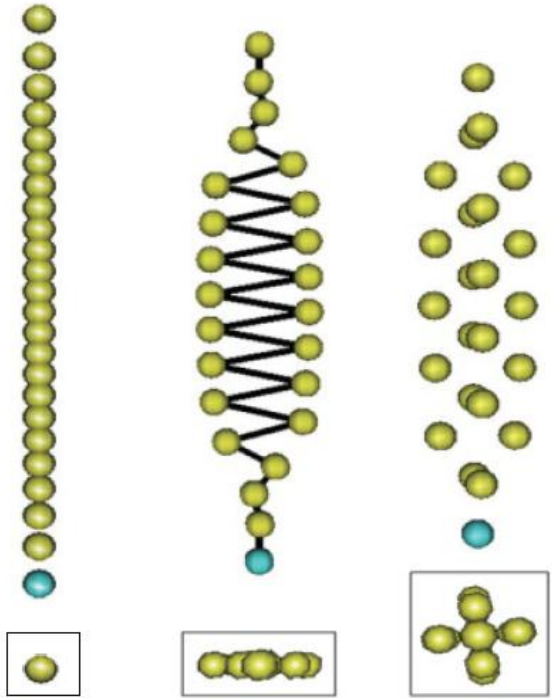

Fig 1. Structural transitions for a complex plasma system (N = 24) for three values of the prolateness parameter $\alpha^{-1}$, (a) 100, (b) 50, and (c) 10 where $\alpha^{-1}$ is defined as in [10]. (See Kamimura and Ishihara [10] for additional details.)

### III. Experimental Methods and Model

The complex plasma experiment described in this paper was carried out in two Gaseous Electronics Conference (GEC) RF reference cells located at the Center for Astrophysics, Space Physics and Engineering Research (CASPER) [22]. Each of these cells contains a grounded upper electrode and a powered lower electrode, capacitively coupled to the system and driven at 13.56 MHz. The distance between upper and lower electrodes is 1.9 cm; a 12.5 mm × 10.5 mm (height × width) glass box was placed on the lower electrode to create the confinement potential needed to establish the initial one-dimensional dust particle chain. All experiments were conducted in Argon plasma at 16 Pa employing RF powers between 2.0 W and 5.0 W. Melamine formaldehyde (MF) particles having a mass density of 1.51 g/cm$^3$ and diameter of 8.89 ± 0.09 μm, as described by the manufacturer [20], were used.

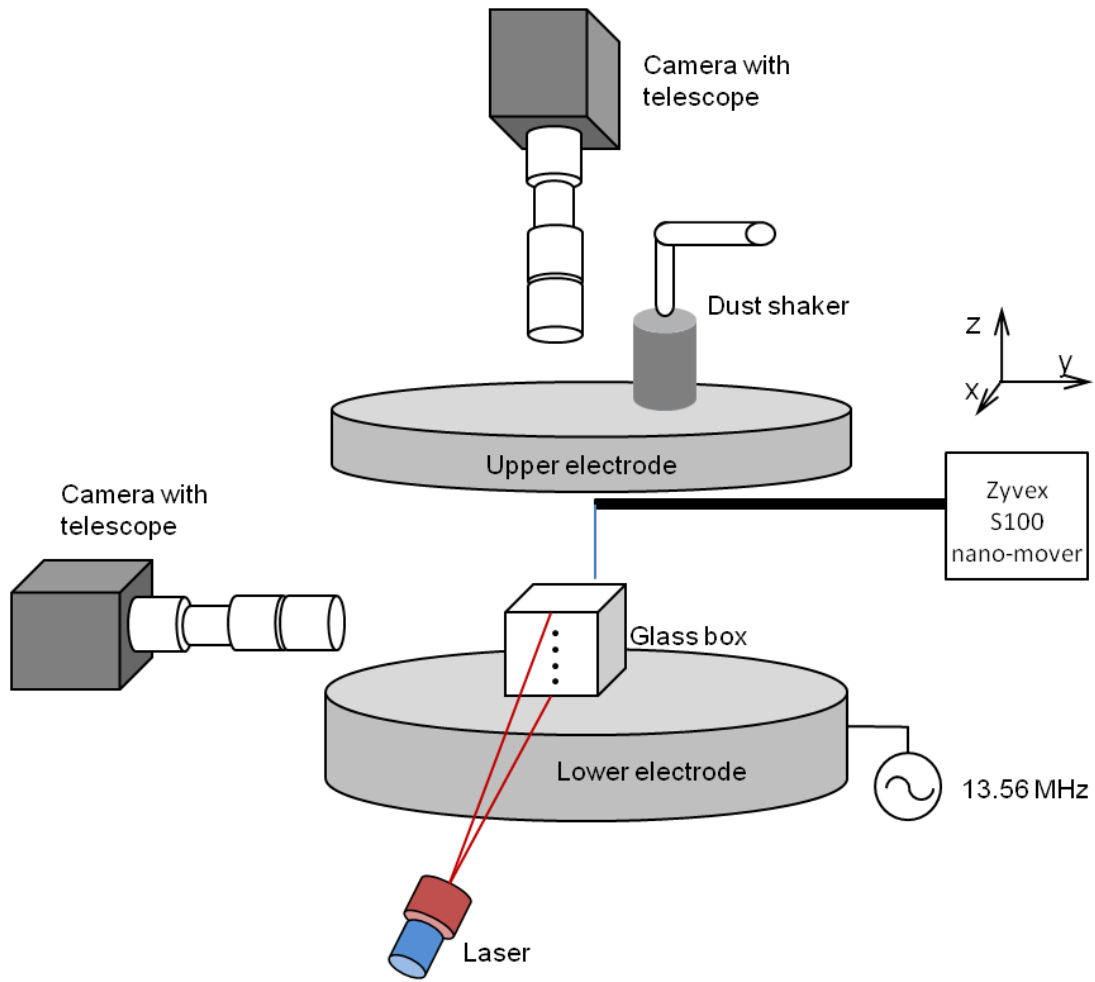

Fig 2. Experimental setup for the experiment described in the text.

Fig 2 provides a schematic of the experimental setup employed. Under the conditions used in this experiment, a vertical region is formed within the glass box exhibiting a constant value of electric force $qE$, where $q$ is the charge on a dust particle and $E$ is the total vertical electric field [21]. This equipotential region (EQP) provides an extended area where the gravitational force and the vertical electric field forces created by the lower electrode and the walls of the box are in balance. For a driving power of 2 W, the measured vertical extent of this region is approximately 5 mm, providing the potential structure necessary for the formation of one-, two- and three-dimensional dust structures.

Varying the RF power to the cell modifies both the EQP and the value of the system confinement parameter, $\alpha$, defined here as the ratio between the horizontal and vertical confinement potential, i.e., $\gamma^2 = (\omega_{0h}/\omega_{0v})^2$, where $\omega_{0h,v}$ are the horizontal and vertical dust resonance frequencies, respectively. Control over these parameters provides the boundary conditions necessary to allow the dust particles to form one-dimensional vertical chains, two-dimensional zigzag structures, and/or three-dimensional helical structures of triangular, quadrangular, pentagonal, hexagonal, and heptagonal symmetries.

For a strongly-coupled dusty plasma made of $N$ identical dust particles of each with charge $Q_d$ and mass $m_d$ confined in a 3D harmonic potential well, $V(r,z) = \frac{1}{2}m_d\omega_{0h}^2 r^2 + \frac{1}{2}m_d\omega_{0v}^2 z^2 = \frac{1}{2}m_d\omega_{0v}^2\left(\gamma^2 r^2 + z^2\right)$, where $\gamma^2 = \omega_{0h}^2/\omega_{0v}^2$ is the system confinement under 3D cylindrical symmetry [14]. The normative technique used for calculating γ is to measure the individual particle resonance frequencies [21]. For the case at hand, a Kepco external DC power supply with its output modulated by a function generator (INSTEK GFG-8210) was used to oscillate the dust particles. This was accomplished by sending a modulated signal to a probe tip attached to a Zyvex S100 nano-manipulator positioned outside the glass box. Since the probe tip was located at an angle to the dust particles (see Fig 3), the dust oscillations generated exhibited both vertical and horizontal components. By scanning the frequency of the probe tip potential from 1 Hz to 15 Hz, the dust particles' horizontal and vertical response spectra were obtained for varying RF powers and then used to calculate the ratio $\gamma^2 = \left(\omega_{0h}^2/\omega_{0v}^2\right)$.

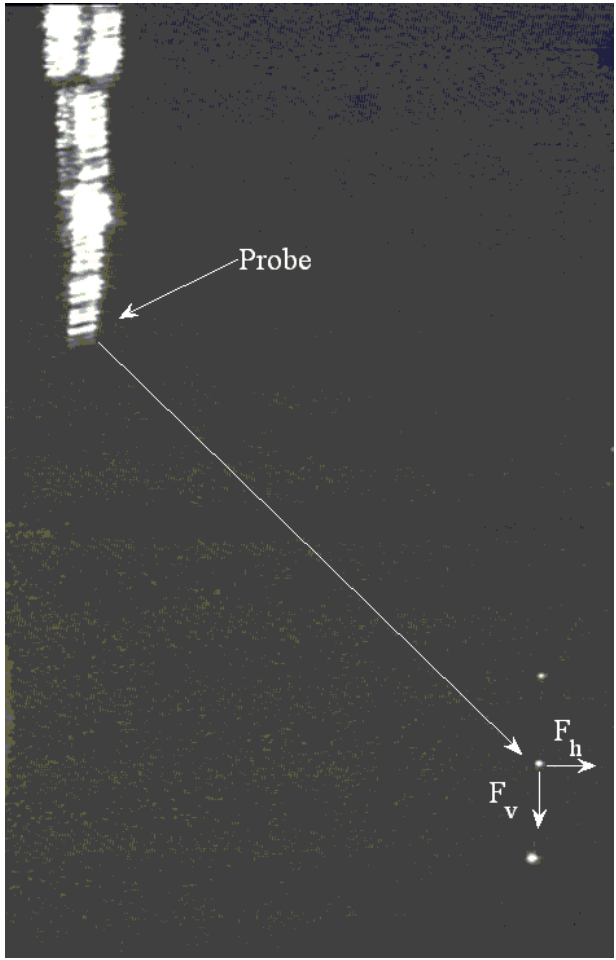

Fig 3. The force exerted on a dust particle from the S100 nano-manipulator probe can be separated into horizontal ($F_h$) and vertical ($F_v$) components.

IV. **Experimental Results**

System power (RF) was initially established at 5 W (500 mV) in order to trap the dust particles within the glass box. At this RF power, the confined dust particles form a turbulent dust cloud consisting of several hundred particles [9]. Lowering the RF power decreases the horizontal radius of the dust cloud while increasing its vertical length; it also controls the total number of dust particles in the box through loss of particles to the lower electrode. In this manner, dust particle bundles of twenty, fourteen, ten and eight particles were prepared for study. For bundles of ten and eight particles, a single vertical chain (i.e., exhibiting one-fold symmetry), a two-dimensional zigzag transition and two chain structure (two-fold symmetry) and three-dimensional three- and four-chain helical structures (exhibiting triangular and quadrangular symmetries) were formed. For bundles of twenty and fourteen particles, three-dimensional four-, six-, seven- and eight-chain helical structures were formed, with pentagonal, hexagonal, and heptagonal symmetries. Figure 4 provides views from above and from the side for each of these structures and denotes the corresponding RF power at which each occurred. The top views shown provide insight into the dimensionality of each structure, while the side view illustrates the symmetry structure.

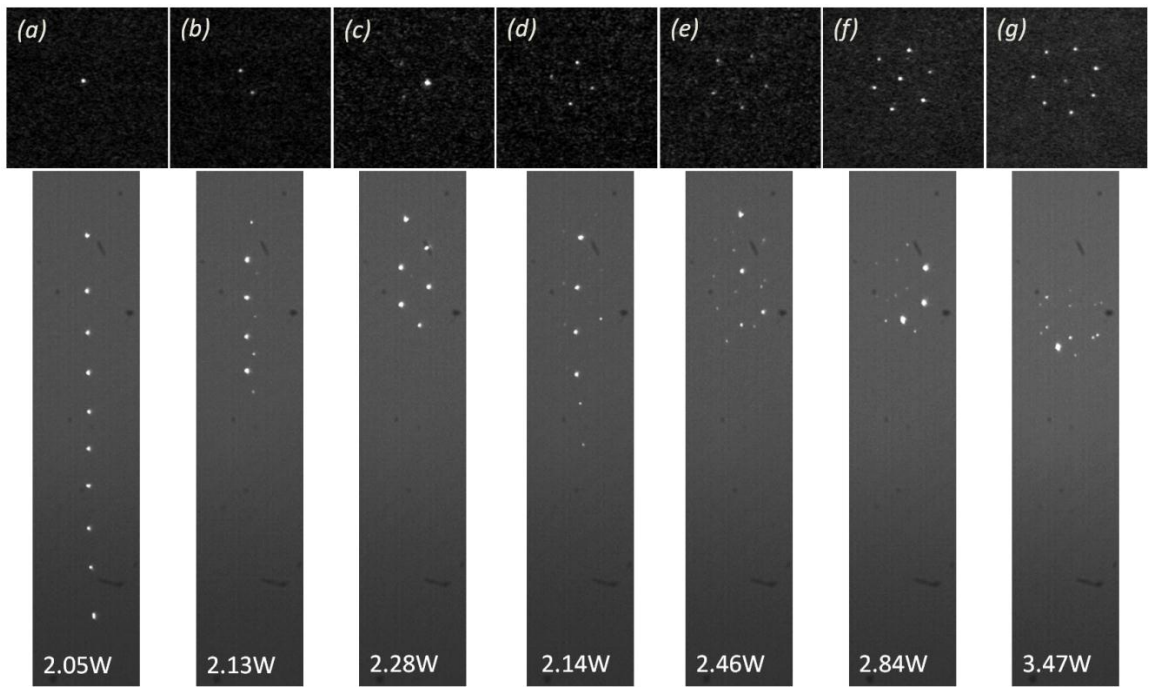

Fig 4. Top view (upper) and side view (lower) of helical structures formed at varying RF powers using the technique described in the text. In all cases, the background pressure is held at 16 Pa. In figures (a) – (c), each structure contains a total of ten particles, while figures (d) – (g) include a total of twenty particles. One- through four-chain structures are shown in (a) through (d), with six through eight chain structures (including the center chain) shown in (e) through (g).

For bundles formed from fourteen or more particles, i.e., in this case containing either fourteen or twenty particles respectively, the minimum symmetry limit is a quadrangular symmetry helical structure for the experimental conditions described (see Fig. 5). Any decrease below 2.14 W in the system's RF power creates a decrease in overall particle number with subsequent transition from a quadrangular (or higher) symmetry structure to a *single* chain structure (see Fig. 5). Additionally, for the $N = 20$ bundles shown above, structural rearrangements between the configurations shown in Fig 4 d-g are both reversible and repeatable, with no noticable hysteresis observed. In all cases, an unstable state exists between these structural rearrangments where particles exihibit short period chaotic movement before transitioning into the next structure. A central vertical chain also forms for six- and seven-fold chains, reducing the asymmetric forces on chain particles.

Each of the above structures exist as stable states, with dust particles exhibiting only slight oscillations about their equilibrium positions, although clockwise and counter-clockwise rotation about the vertical axis was observed during adjustments in the RF power. As can be seen in Fig 5, transitions between structures occur in a stepwise manner and depend upon both the RF power and the total number of particles. It is also interesting to note that larger numbers of dust particles form multiple chain structures at lower RF powers. For example, a four-chain helical structure can form from twenty particles, fourteen particles, ten particles or eight particles, at RF powers of 2.13 W, 2.16 W, 2.29 W and 2.68 W, respectively.

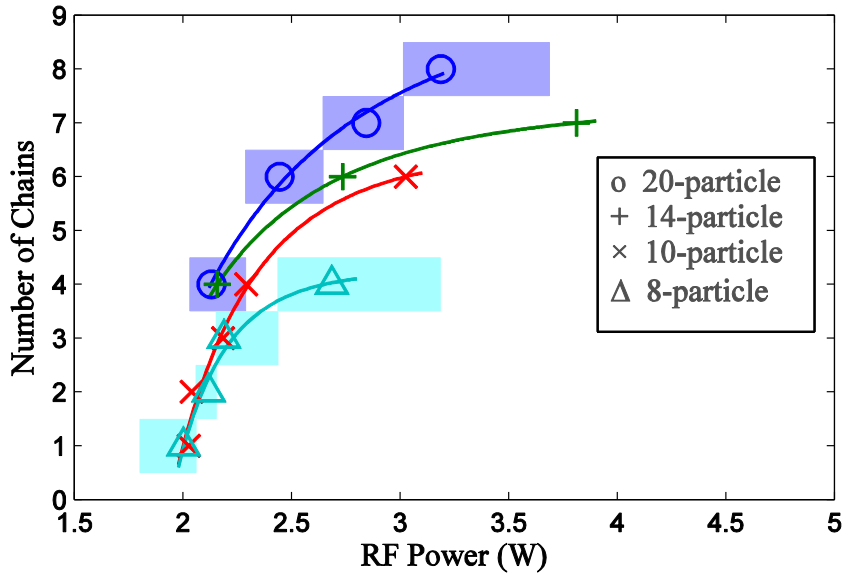

Fig 5. Structural arrangement as a function of system RF power. The number of distinct chains depends on both the number of dust particles and the RF power. The symbols shown represent the RF power at which a given structural arrangement exhibited its most stable state (i.e., minimum rotation). The shaded bars shown for the 20- and 8-particle chains indicate the range of RF power over which a given structure can exist, indicating the step-wise nature of the transitions. Fit lines are included to guide the eye.

Defining $R_h$ as the horizontal radius of the helix, $\Delta$ as the helical vertical period, and $R_d$ as the interparticle separation distance between nearest neigbors (see Figure 6), the relationship between the horizontal and vertical dimensions for a given helical structure can be examined. As shown, the nearest neighbor separation distance, $R_d$, represents a measure of the compactness of the structure while the ratio $R_h/\Delta$ provides the prolateness as defined in [10].

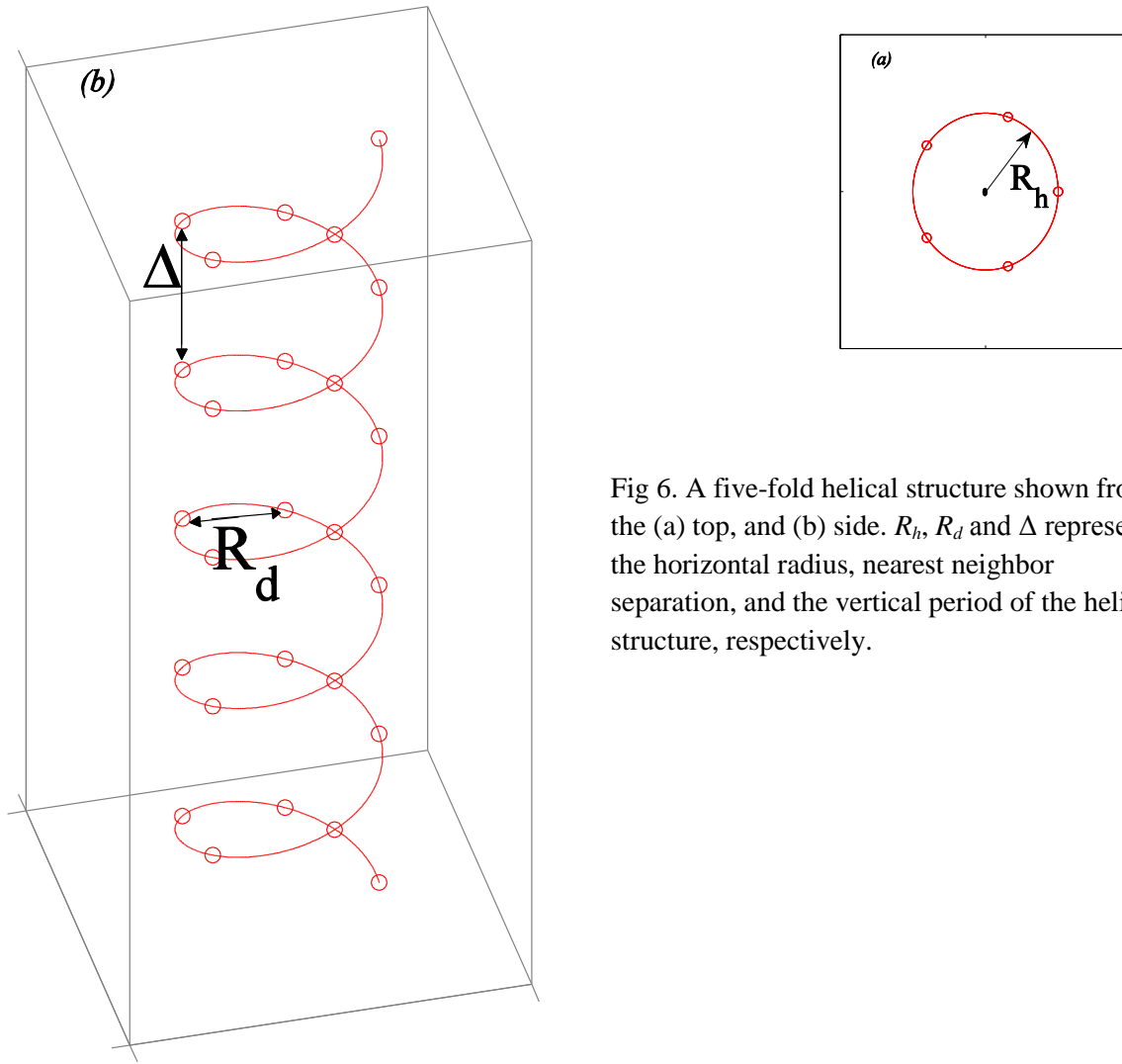

Fig 6. A five-fold helical structure shown from the (a) top, and (b) side. $R_h$, $R_d$ and $\Delta$ represent the horizontal radius, nearest neighbor separation, and the vertical period of the helical structure, respectively.

For the two- through eight-chain helical structures examined above, the nearest neighbor separation distance $R_d$ is a function of both the total particle number and the number of chains. As the total particle number or the number of chains increases, $R_d$ decreases (Fig 7) compacting the helical structure.

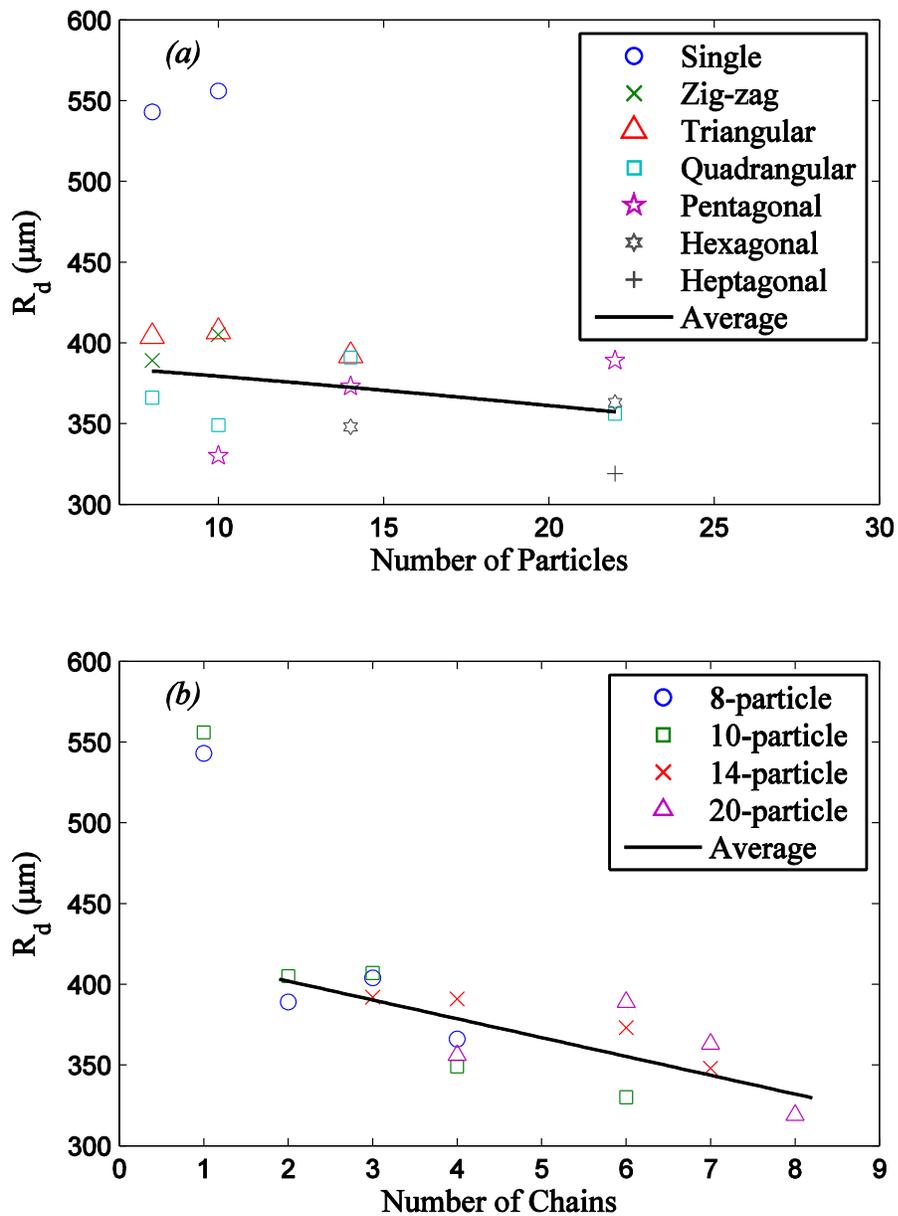

Fig 7. (a) Interparticle separation distance, $R_d$, as a function of total particle number, $N$. For two- to eight-chain structures, the average interparticle separation distance decreases linearly as the total number of particles increases. (b) $R_d$ as a function of the number of chains for varying values of $N$. As the number of chains increases, $R_d$ decreases (a linear fit to the average is indicated by the solid line).

The helical 'prolateness,' defined as $R_h/\Delta$, is shown in Fig 8. As expected, this value is bounded between zero and one, tending to zero as the number of chains decreases to one, and to one as the number of chains in the cluster increases.

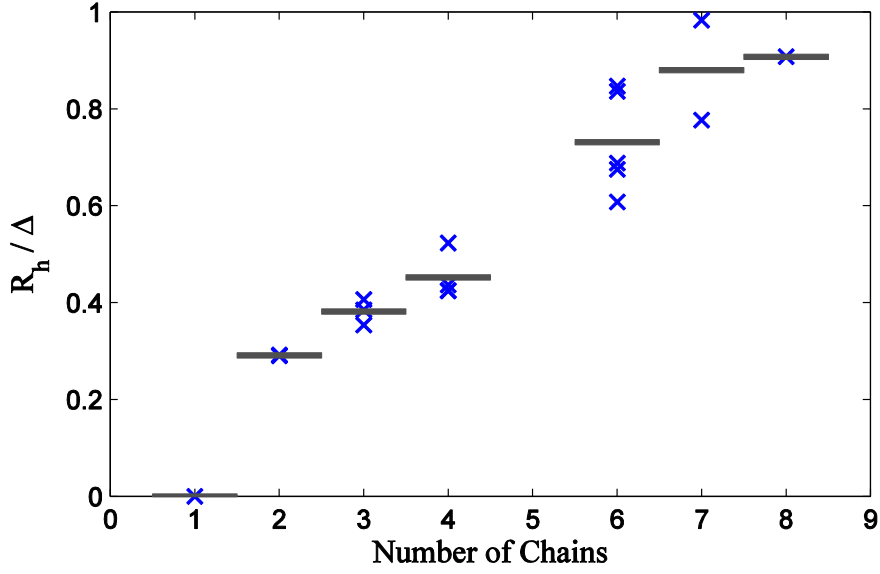

Fig 8. Helical 'prolateness', defined as $R_h/\Delta$, as a function of the number of chains. The bars indicate the average of these measurements.

Finally the system confinement, $\gamma^2 = (\omega_{0h}/\omega_{0v})^2$, was measured for the experimental conditions employed in this study. This was accomplished by establishing oscillations about the center of mass for a vertically aligned two-particle chain and measuring the resulting horizontal and vertical response spectra. Resonance frequencies for both vertical and horizontal oscillations were provided through a frequency sweep of the probe potential at various RF powers.

The overall frequency response is related to the dust particle charge, $\omega_{0h,v}^2 = \rho_{h,v} Q_d / \varepsilon_0 m_d$, where $\rho_{h,v}$ is the charge density of the undisturbed plasma sheath in the horizontal or vertical direction [17, 18]. The plasma charge density in each direction can be found through approximating the variation in the electric field using a Taylor series expansion:

$$Q_d E = Q_d \left( E_0 + E' \Delta x + \frac{1}{2} E'' (\Delta x)^2 + ... \right) \quad (1)$$

The electric field is related to the plasma charge density through Poisson's equation, which when separated into horizontal and vertical components yields,

$$\nabla^2 \varphi = \frac{\rho}{\varepsilon_0} \Rightarrow -\nabla \cdot E = \frac{\rho}{\varepsilon_0} \Rightarrow \left( \frac{\partial E_x}{\partial x} + \frac{\partial E_y}{\partial y} \right) = -\frac{\rho}{\varepsilon_0} \quad (2)$$

$$\frac{\partial E_x}{\partial x} = -\frac{\rho_x}{\varepsilon_0}$$
$$\frac{\partial E_y}{\partial y} = -\frac{\rho_y}{\varepsilon_0} \quad (3)$$

where $\rho_x + \rho_y = \rho$. Substituting (3) into (1) and keeping only first order terms yields the relations

$$Q_d E_x = Q_d E_{0x} + \left(-\frac{\rho_x Q_d}{\varepsilon_0}\right)\Delta x$$
$$Q_d E_y = Q_d E_{0y} + \left(-\frac{\rho_y Q_d}{\varepsilon_0}\right)\Delta y \tag{4}$$

In the above, $Q_d E_{0y}$ is balanced by the gravitational force, while $Q_d E_{0x} = 0$, given that at equilibrium the net horizontal confinement force is zero. Therefore, for small displacements,

$$Q_d \Delta E_x = -k_x \Delta x$$
$$Q_d \Delta E_y = -k_y \Delta y \tag{5}$$

where $k_{x,y} = \rho_{x,y} Q_d / \varepsilon_0$.

The lowest power for which particles could be suspended under the operating parameters employed was found to be 2.05 W. Since both $\rho_{x,y}$ and $Q_d$ are functions of the RF power and $k_{x,y}$ is related to the horizontal or vertical confinement strength, $k_{x,y}$ should increase when the RF power is increased. Experimental results verify this (see the following section) and also show that $k_y$ increases faster than $k_x$.

As previously mentioned, the confinement parameter, $\gamma^2 = (\omega_{0h}/\omega_{0v})^2$, is a measure of the system confinement ratio in the radial and vertical directions and is calculated by measuring the resonant frequencies in the horizontal and vertical directions. Representative response spectra obtained in this manner for horizontal and vertical oscillations at RF powers of 2.08 W and 3.37 W are shown in Fig 9(a) and 9(b). Fig 9(c) shows the resulting resonance frequency distribution as a function of the RF power while Fig 9(d) shows the confinement parameter $\gamma^2 = (\omega_{0h}/\omega_{0v})^2$ calculated from the data in Fig 9(c), plotted as a function of the RF power. In agreement with the analysis above, the resonance frequency decreases as the RF power decreases, with the horizontal resonance frequency decreasing at a slower rate than the vertical.

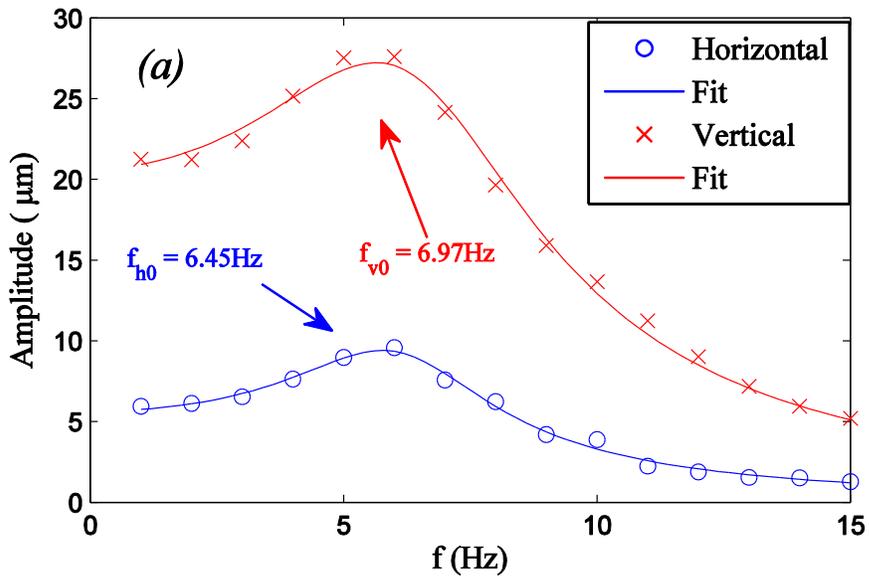

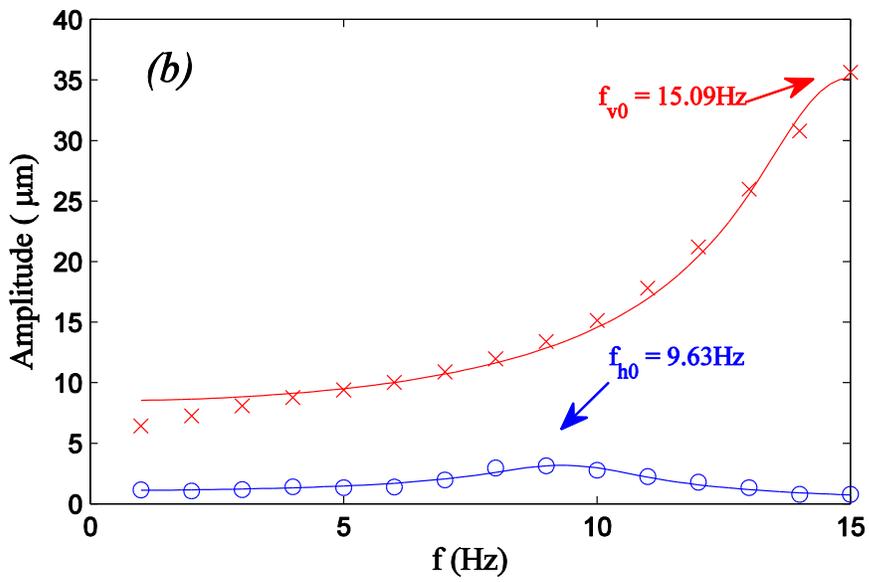

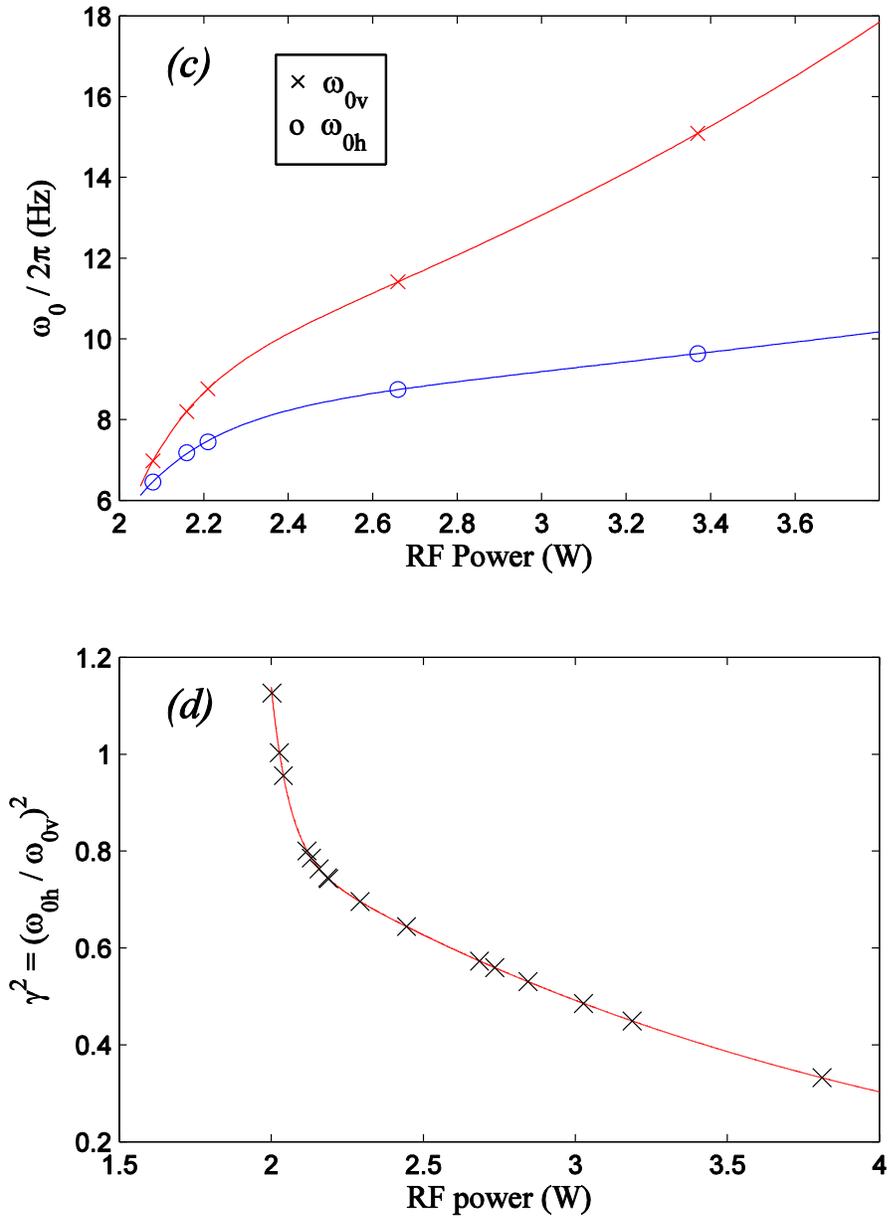

Fig 9. Horizontal and vertical resonance frequency curves for system RF powers of (a) 2.08 W and (b) 3.37 W. Data points shown indicate driving frequencies, while lines provide a damped linear oscillation fit to this data. (c) Resonance frequency distribution as a function of RF power. The solid lines are polynomial fits to guide the eye. (d) Confinement parameter $\gamma^2$, calculated from the data shown in (c). The solid line serves to guide the eye.

Measured system confinement $\gamma^2 = \left(\omega_{0h}/\omega_{0v}\right)^2$ as a function of particle number for the helical structures observed is shown in Fig 10. As shown, the total number of particles $N$ for a given structure decreases as $\gamma^2$ decreases (with the exception of the single vertical chain), in agreement with results from [10].

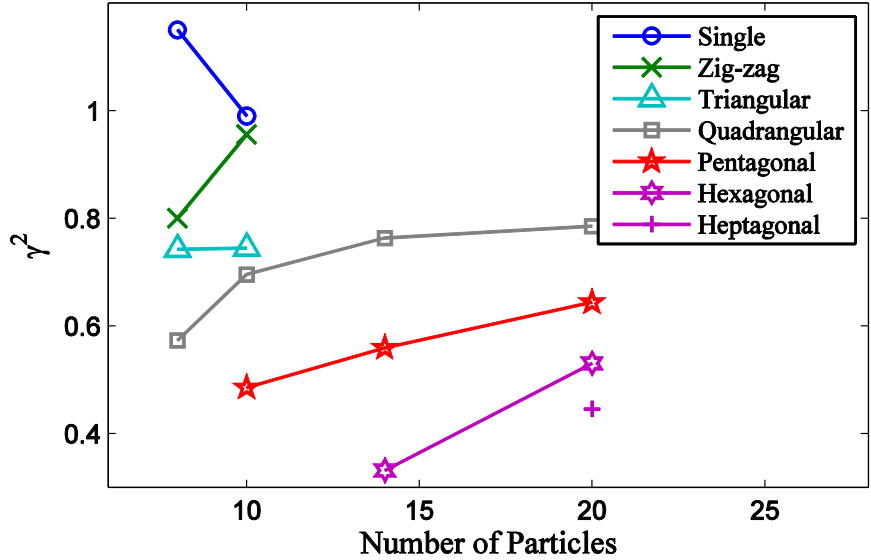

Fig 10. System confinement ($\gamma^2$) versus total number of particles in a stable structure. As the confinement decreases, the total number of particles contained within a given structural arrangement decreases, in agreement with Kamimura and Ishihara (2012) [10]. The lines connecting the data points are meant to guide the eye.

For a conserved number of dust particles, critical values of $\gamma^2$ exist marking the points at which instabilities lead to the emergence of transitions between consecutive stable structures. For example, for the 10-particle cluster above, the transition from a pentagonal to quadrangular symmetry occurs at $\gamma^2 = 0.6$, from a quadrangular to triangular symmetry at $\gamma^2 = 0.7$, and from a triangular to zig-zag symmetry at $\gamma^2 = 0.85$. The resulting boundary conditions are therefore (1) for a single chain $\gamma^2 \geq 1$, (2) for zig-zag structures (2-chain) $0.8 < \gamma^2 < 1$, and for (3) helical structures (triangular to heptangonal) $\gamma^2 < 0.8$, and thus the critical value for helical structure formation is $\gamma^2 = 0.8$.

## V. Discussion and Conclusions

In this work vertically aligned, charged dust particles levitated within a glass box placed on the lower, powered electrode of a RF GEC cell were examined. The overall plasma was assumed to be charge neutral; in other words, free electrons and electrons on the charged dust particles were assumed to balance the number of ions. The system was also assumed to be in a state far from thermal equilibrium and large in all dimensions as compared to the Debye-Huckel shielding length. Three-dimensional confinement within the box was provided by the horizontal and vertical electric fields produced by the box walls, the vertical gravitational force and the sheath electric field. Therefore, the charged dust particles were confined through a combination of gravity, the effective potential produced by the dust-plasma interaction, and the confinement provided by the glass box on the lower electrode.

Self-organized formation of one-dimensional vertical chains, two-dimensional zigzag structures, and three-dimensional helical structures of triangular, quadrangular, pentagonal, hexagonal, and heptangonal symmetries were examined in detail. Control over various system operating parameters was shown to determine overall dust particle structural symmetry; employing this technique, the evolution from a one-dimensional chain structure, through a zigzag transition to a two-dimensional, spindle-like structure and then to various three-dimensional, helical structures exhibiting multiple symmetries was observed.

Varying the RF power of the system produced a zigzag transition between vertically aligned one- and two-dimensional structures. This was shown to be dependent upon both the total particle number and the change in anisotropy of the confinement (i.e., RF power). These results are in partial agreement with those reported by Melzer [13] which showed experimentally that the horizontal zigzag transition was dependent upon both particle number and change in anisotropy of the confinement. However, Melzer did not find the horizontal zigzag transition to be dependent upon changes in plasma power, since system confinement was applied in his case via a horizontal, rectangular barrier; this in turn provided a confinement parameter, $\alpha^2 = (\omega_{0x}/\omega_{0y})^2$, primarily defined by the geometric ratio of the barrier. In the experiment described here, overall system geometry remains fixed with horizontal (or radial) confinement created by the surface charge density on the walls of the glass box and related to the plasma screening length. As such, it is much more sensitive to changes in RF power than is the vertical confinement, which arises primarily from the sheath potential.

Decreasing the anisotropy of the confinement ($\gamma^2$) leads to a corresponding decrease in the prolateness parameter ($\alpha^{-1}$) as defined in Kamimura and Ishihara [10]. In this case, trapped dust particles were shown to undergo a series of phase transitions between a one-dimensional, single chain structure, a two-dimensional structure and a series of three-dimensional dust clusters, all of which consist of multiple dust particle chains exhibiting spindle-like/helical symmetry. The results identified theoretically in [10] were based on configurations of minimum energy (CME) with transitions from single to double to quadruple to sextuple chains identified. The experimental results shown here are in good agreement with this data (compare Fig 10 in this paper with Figure 2 in Reference 10); however additional structures in one-, two- and three-dimensions with triangular- to heptagonal-symmetry have been shown to exist experimentally.

Dust cluster structural symmetry was shown to be dependent upon system confinement $\gamma^2 = (\omega_{0h}/\omega_{0v})^2$, with transitions between symmetries occurring as (1) one- to two-chain zigzag structures, (2) two-chain zigzag to triangular structures, (3) triangular to quadrangular structures, (4) quadrangular to pentagonal structures, (5) pentagonal to hexagonal structures, and (6) hexagonal to heptagonal structures. Critical values of γ were determined for each of these (see Fig 10) and shown to be a function of both the RF power and *N*. For clusters having a fixed number of particles, the RF power at which these transitions occur is repeatable and exhibits no observable hysteresis.

In addition to structural symmetry, the confinement ($\gamma^2$) also establishes the available energy phase space; this in turn determines the overall number of chains that can form under a given set of operating conditions. This result is directly related to the overall vertical extension of the 'trap' which controls the initial number of particles that can be captured and, subsequently, the total number of particles within a given vertical chain. For the operating conditions employed here, at RF powers < 2.14 W, a maximum of ten particles within a single vertical chain was found to exist. Increasing the RF power altered the anisotropy of the confinement $\gamma^2$ such that the overall vertical length of the cluster decreased while its horizontal (radial) extent increased. This is due to the fact that although both $\omega_{0h}^2$ and $\omega_{0v}^2$ increase as the RF power increases, $\omega_{0v}^2$ increases faster than $\omega_{0h}^2$. Therefore as the system RF power increases, the dust cluster shrinks in the vertical direction while expanding in the horizontal (radial) direction.

At higher RF powers (> 2.14 W), the phase space can maintain more than ten particles (for example, the fourteen and twenty particles discussed above) allowing the formation of higher order symmetry structures comprised of increased numbers of chains. For bundles having fixed numbers of particles greater than fourteen, structural rearrangements between these configurations are both reversible and repeatable, with no noticable hysteresis. However once the RF power is reduced below 2.14 W, $\omega_{0v}$ decreases rapidly causing $\gamma^2$ to increase rapidly. This results in a series of stepwise structural transitions from a four-chain (or higher) structure directly to a single chain structure. For the operating conditions examined by this study, the minimum symmetry structure for clusters having more than ten particles was found to be a four-fold helical structure.

It was also determined that the ion drag force contributes only minimally to the topology of the trap. This appears to be in agreement the results previously published by Arp, et al. [15], although this is not meant to imply that there is no ion drag force within the box or that this result is true under all circumstances. A strong asymmetry of the particle coupling in the vertical direction was observed which cannot be explained by a pure ion wake potential alone [9]. This appears to be in agreement with the theory of an effective force arising from the confinement (perhaps due to a variable charge) as presented by Carstensen, at al. [18]. Both of these will be addressed in a future publication.

Finally, it is interesting to note that helical structures similar to those described above have been observed in both non-neutral and/or one-component plasmas. As such, dusty plasmas continue to prove themselves excellent 'analogues' for the examination of the micro-dynamics of strongly coupled Coulomb systems.